\documentclass[prb,twocolumn,amsmath,amssymb,citeautoscript,showkeys,floatfix,superscriptaddress]{revtex4-2}   

\usepackage{chemformula} 
\usepackage[T1]{fontenc} 
\usepackage{siunitx}
\usepackage{amssymb}
\usepackage{amsmath}
\usepackage{graphicx}
\usepackage{multirow}
\usepackage{booktabs}
\usepackage{footnote}
\usepackage[flushleft]{threeparttable}
\usepackage{comment}
\usepackage[normalem]{ulem}

\usepackage{hyperref}
\hypersetup{
    colorlinks,%
    citecolor=blue,%
    linkcolor=blue,%
    urlcolor=blue
}

\newcommand{\orcid}[1]{\href{https://orcid.org/#1}{\includegraphics[width=8pt]{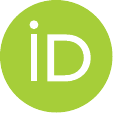}}}

\usepackage{filecontents}

\begin{document}

\title{The deep-acceptor nature of the chalcogen vacancies in 2D transition-metal dichalcogenides}

\author{Shoaib Khalid\orcid{0000-0003-3806-3827}} 
\affiliation{Computational Sciences Department, Princeton Plasma Physics Laboratory, Princeton, NJ 08540, USA}
\author{Bharat Medasani\orcid{0000-0002-2073-4162}}
\affiliation{Computational Sciences Department, Princeton Plasma Physics Laboratory, Princeton, NJ 08540, USA}
\author{John L. Lyons\orcid{0000-0001-8023-3055}}
\affiliation{Center for Computational Materials Science,   U.S. Naval Research Laboratory, Washington, D.C. 20375, USA}
\author{Darshana Wickramaratne\orcid{0000-0002-1663-1507}}
\affiliation{Center for Computational Materials Science,   U.S. Naval Research Laboratory, Washington, D.C. 20375, USA}
\author{Anderson Janotti\orcid{0000-0002-0358-2101}}
\affiliation{Department of Materials Science and Engineering, University of Delaware, Newark, DE 19716, USA}

\begin{abstract}

Chalcogen vacancies in the semiconducting monolayer transition-metal dichalcogenides (TMDs)
have frequently been invoked to explain a wide range of phenomena, including both unintentional p-type
and n-type conductivity, as well as sub-band gap defect levels measured via tunneling or
optical spectroscopy. These conflicting interpretations of the deep versus shallow nature of the chalcogen vacancies are due in part to shortcomings in prior first-principles calculations of defects in the semiconducting two-dimensional (2D) TMDs that have been used to explain experimental observations. Here we report results of hybrid density functional calculations for the chalcogen vacancy in a series of monolayer TMDs, correctly referencing the thermodynamic charge transition levels to the fundamental band gap (as opposed to the optical band gap). We find that the chalcogen vacancies are deep acceptors and cannot lead to n-type or p-type conductivity. Both the (0/$-1$) and ($-$1/$-$2) transition levels occur in the gap, leading to paramagnetic charge states S=1/2 and S=1, respectively, in a collinear-spin representation. We discuss trends in terms of the band alignments between the TMDs, which can serve as a guide to future experimental studies of vacancy behavior.
\end{abstract}


\maketitle

Layered transition-metal dichalcogenides with formula unit MX$_2$ (where M is a transition-metal, and X represents a chalcogen atom) are of technological interest due to their unique mechanical\cite{castellanos2012elastic,Peng_2023}, electrical\cite{radisavljevic2011single,lopez2013ultrasensitive}, and optical properties~\cite{chhowalla2013chemistry,late2013sensing}.  They
have been studied for a variety of applications, including optoelectronics \cite{radisavljevic2011single,lopez2013ultrasensitive}, sensors \cite{late2013sensing}, field-effect transistors \cite{radisavljevic2011single}, heterostructure junctions \cite{georgiou2013vertical}, photovoltaics\cite{bernardi2013extraordinary}, photodetectors\cite{choi2012high}, and single-photon emitters\cite{koperski2015single,kern2016nanoscale}.
Those TMDs that are semiconductors have bands gaps and related properties that critically depend on the number of layers \cite{radisavljevic2011single,fai2010atom,di2012coupled,mak2013tightly,bertolazzi2011stretching,castellanos2012elastic}. Intrinsic defects are often invoked to explain many of the interesting properties of TMDs, and their impact is expected to be much higher in the monolayer than in their bulk counterpart \cite{stanford2016focused,Tomer2022tunable,lee2018tuning,Ghorbani2013defect,korn2011low,Abramson2018defect,liang2021defect}.
Some defects are created unintentionally during growth, and others can be intentionally introduced in post-growth processes, such as exfoliation, annealing, or irradiation \cite{liang2021defect}. One of the most commonly invoked defects for interpreting a wide range of electrical and optical properties \cite{shen2022healing,zhou2013intrinsic,vancso2016intrinsic,guo2015chalcogen,yuan2014effect} is the chalcogen vacancy, and its influence on the electronic and optical properties of TMDs has been the subject of intense debate \cite{shen2022healing,zhou2013intrinsic,vancso2016intrinsic,guo2015chalcogen,yuan2014effect,lien2019electrical,tang2021effects}. For instance, in single-layer MoS$_2$, MoSe$_2$, and WSe$_2$, it is commonly argued that the chalcogen vacancy is a shallow donor that gives rise to the often observed n-type conductivity \cite{shen2022healing,lee2018tuning,Ghorbani2013defect,tosun2016air}. Other studies have invoked chalcogen vacancies to explain the origin of sub-bandgap photoluminescence peaks that exhibit single-photon emitter characteristics, implying that these vacancies are deep-level centers \cite{mitterreiter2021role}.

\begin{figure*}
\includegraphics[width=12 cm]{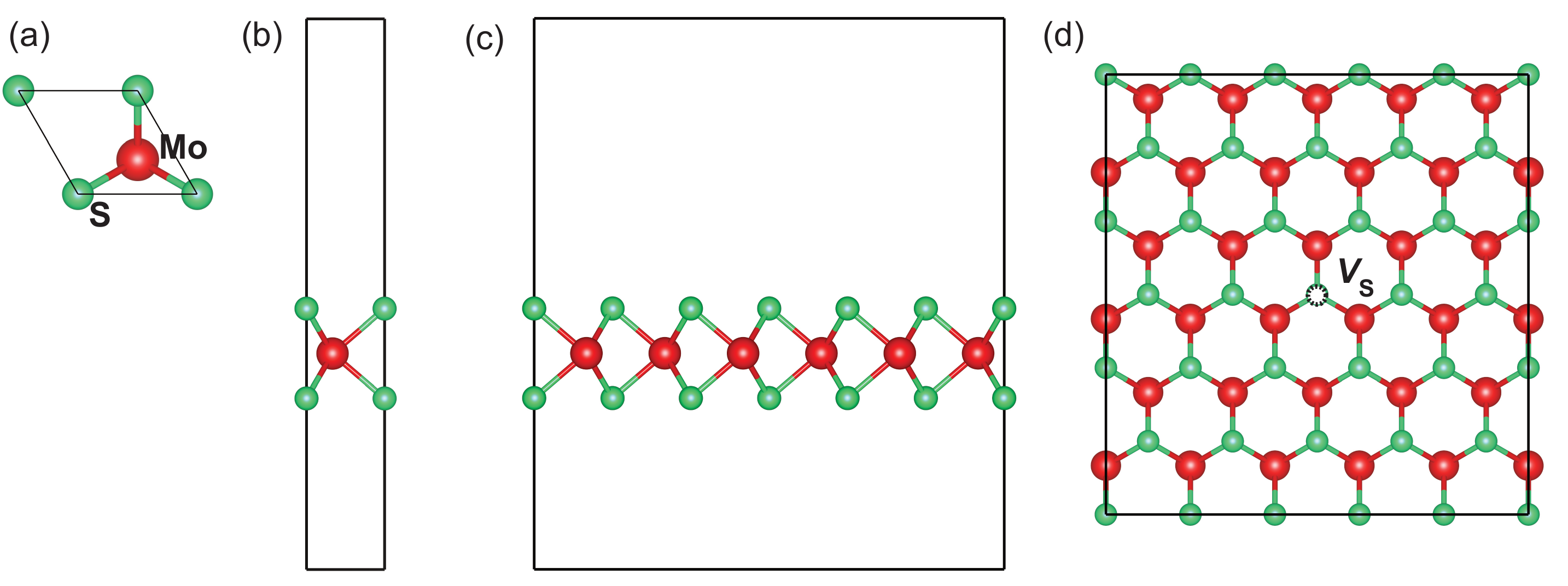}
\caption{Primitive cell of monolayer TMDs: (a) top view and (b) side view showing a vacuum thickness of $\sim$19 \AA. The supercell used for the defect calculations, containing 90 atoms: (c) side view and (d) top view.}
\label{fig1}
\end{figure*}

First-principles calculations are instrumental in elucidating the electronic properties of defects in semiconductors, determining whether they are shallow or deep centers, and if they exhibit donor or acceptor character\cite{Freysoldt2014}. In fact, there have been several theoretical reports of defects in these semiconducting monolayer TMDs \cite{komsa2015native,abhishek2019origin,shang2018elimination,Felipe2022,qiu2022first}. One principal issue with these prior calculations is that the defect thermodynamic transition levels were calculated with respect to the optical band gap, instead of properly referencing to the fundamental band gap. Semi-local functionals coincidentally lead to calculated band gaps that are close to the optical band gaps, which are significantly lower (by 0.6 to 0.9 eV) than the fundamental electronic band gap due to large exciton binding energies \cite{guo2016band,Lambrecht2012,Robert2016}. Attempts to correct for the underestimated fundamental band gap using more advanced functionals \cite{tan2020first} used values of the band gap that are still lower than the experimentally established
values. Such calculations fail to provide quantitatively accurate predictions of defect levels and formation energies, precluding a reliable comparison with experimental data. This failure is due to a multitude of factors, including using the incorrect value of the fundamental band gap to
determine defect-level positions, the inability to describe the position of band edges correctly, and the inability of semilocal functionals to describe carrier localization\cite{Freysoldt2014}.

The second issue is the convergence of the formation energy of charged defects with respect to the "vacuum thickness". Since the lattice periodicity in plane-wave DFT calculations is three-dimensional, calculations for charged defects using periodically repeated slab geometries in a 3D periodic lattice lead to significant and uncontrolled errors if not properly addressed. Only recently have formal and systematic approaches to solving this problem been successfully addressed and implemented \cite{silva2021}. In this work, we address these two major problems and present results for the chalcogen vacancy formation energies and thermodynamic transition levels for the series of monolayer TMDs comprised of MoS$_2$, MoSe$_2$, MoTe$_2$, WS$_2$, WSe$_2$ and WTe$_2$ in the 1H semiconducting crystal structure. We discuss the results in terms of band alignments between the different TMDs by properly taking fundamental band gaps into account.

\begin{table}
\begin{center}
\caption{Lattice parameter of monolayer transition-metal dichalcogenides (TMDs) in the 1H crystal structure calculated using the SCAN and HSE$\alpha$ functionals. The experimental data are taken from \cite{huang2015bandgap,gusakova2017electronic,chen2017chemical,huang2015large,yu2017molecular}.}
\begin{threeparttable}
\setlength{\tabcolsep}{8pt} 
\renewcommand{\arraystretch}{1.5} 
\begin{tabular}{lccc}
  \toprule\toprule
  \multirow{2}{*}{\raisebox{-\heavyrulewidth}{Material}} & \multicolumn{3}{c}{Lattice parameter ({\AA}) } \\
  \cmidrule(lr){2-4}
  & SCAN & HSE$\alpha$ & Exp. \\
  \midrule
  MoS$_2$ & 3.167 & 3.121 & 3.160\cite{gusakova2017electronic}, $3.20\pm0.1$ \cite{huang2015bandgap}  \\
  MoSe$_2$ & 3.292 & 3.241 & 3.288\cite{gusakova2017electronic}, $3.3\pm0.1$\cite{chen2017chemical}    \\
  MoTe$_2$ & 3.503 & 3.454 & 3.520 \cite{yu2017molecular}   \\
  WS$_2$ & 3.160 & 3.115 &  3.153\cite{gusakova2017electronic}    \\
  WSe$_2$ & 3.285 & 3.230 & 3.290\cite{huang2015large}, 3.280\cite{gusakova2017electronic}    \\
  WTe$_2$ & 3.501 & 3.446 & -   \\
   \bottomrule\bottomrule
\end{tabular}
\label{table:lattice_parameter}
\end{threeparttable}
\end{center}
\end{table}

Our calculations are based on the density functional theory (DFT) \cite{hohenberg1964inhomogeneous,kohn1965self} with the projector augmented wave (PAW) potentials \cite{blochl1994projector} as implemented in the VASP code \cite{kresse1993ab,kresse1994ab}.
We used both the screened hybrid functional of Heyd-Scuseria-Ernzerhof (HSE$\alpha$) \cite{heyd2003hybrid,heyd2006erratum}, and also the meta-GGA strongly constrained and appropriately normed (SCAN) functional \cite{Perdew2015} for comparison.  In the screened HSE hybrid functional, the exchange potential is divided into short-range and long-range parts. The generalized gradient approximation (GGA) exchange potential \cite{perdew1996generalized} is used for the long-range part, whereas in the short-range, the Fock exchange is mixed with the GGA exchange with a mixing parameter $\alpha$. In this work, the value of $\alpha$ is set to 0.40 (40\% of Fock exchange) in order to correctly describe the fundamental band gaps of monolayer TMDs, which are in good agreement with results from quasi-particle GW calculations \cite{Kim2021,Lambrecht2012,Ugeda2014,Robert2016,gusakova2017electronic}. We note that the effects of spin-orbit coupling (SOC) are included. 

The equilibrium lattice parameters for the monolayer TMDs were calculated using both the SCAN \cite{Perdew2015} and HSE$\alpha$ functionals for the three-atom primitive cell with a vacuum thickness of 19 {\AA} [Fig.~\ref{fig1}(a) and (b)] and a 6$\times$6$\times$1 $\Gamma$-centered mesh of $k$ points for integrating over the Brillouin zone.  All the calculations are done with a kinetic energy cutoff of 400 eV for the plane-wave basis set. For the defect calculations, we used a supercell with orthogonal lattice vectors containing 90 atoms as shown in Fig.~\ref{fig1}(c) and (d), and a 4$\times$4$\times$1 $\Gamma$-centered mesh of $k$ points. All atoms in the supercell are allowed to relax, to a maximum force of 0.005 eV/\AA. To avoid the spurious effects of having a neutralizing background homogeneously distributed over the whole supercell not matching the charge distribution of the charged defect with wavefunction localized on the 2D material, we employed the recently developed self-consistent charge-state correction (SCPC) \cite{silva2021}, as implemented in the VASP code. The results using the SCPC method show that the defect formation energy for a negatively charged vacancy in MoS$_2$ changes by less than 0.1 eV when varying the vacuum thickness from 16 \AA~to 30 \AA.

\begin{figure*}
\includegraphics[width=14 cm]{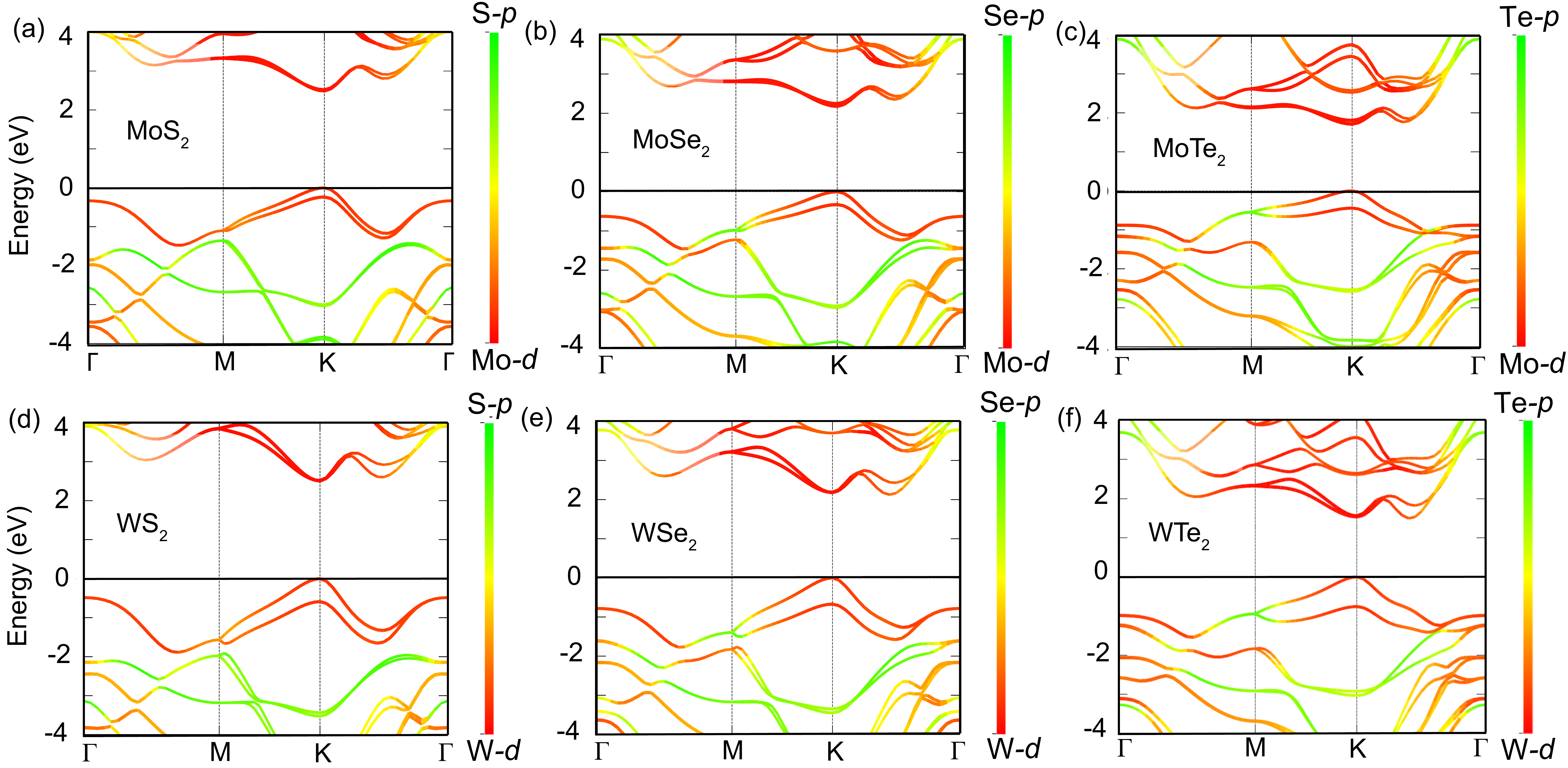}
\centering
\caption{Orbital-resolved electronic band structures of (a) MoS$_2$, (b) MoSe$_2$, (c) MoTe$_2$, (d) WS$_2$ (e) WSe$_2$ and (f) WTe$_2$ using the HSE$\alpha$ hybrid functional and including spin-orbit coupling. The valence-band maximum (VBM) in each case is set to zero.}
\label{fig2}
\end{figure*}

The calculated equilibrium lattice parameters of the TMDs considered in the present work are listed in  Table~\ref{table:lattice_parameter}. The results using SCAN and HSE$\alpha$ are within 1\%-1.5\% of the experimental data \cite{huang2015bandgap,gusakova2017electronic,chen2017chemical,huang2015large,yu2017molecular}.  The lattice parameters of TMDs weakly depend on the metal atom (i.e., Mo vs W), but strongly depend on the chalcogen atom; the lattice parameter significantly increases by going from the lighter sulfur S atom to the heavier Te atom, whereas the lattice parameters change by less than 0.5\% between MoS$_2$ and WS$_2$. For the heavier chalcogen atoms, the difference is slightly larger but still less than 1.5\%.  Therefore, the variations in structural parameters across this family of TMDs are mostly governed by the chalcogen atom. The small lattice mismatch between the common chalcogen MoX$_2$ and WX$_2$ suggests that these materials are ideal for MoX$_2$/WX$_2$ heterostructures with either interface defined by the van der Waals gap or in-plane interfaces, with minimal strain or structural defects \cite{tongay_tuning_2014}. We expect that the small mismatch could also lead to moir\'e patterns with large-scale periodicity, and interesting electronic structure features are likely to manifest.

\begin{table}
\begin{center}
\caption{Fundamental band gap of monolayer TMDs calculated using SCAN and HSE$\alpha$. Previous results from quasi-particle GW calculations and experimental results are taken from \cite{Kim2021,Lambrecht2012,Robert2016,gusakova2017electronic} and \cite{zhang_probing_2015,yang2015robust,gusakova2017electronic,Schuler2019}, respectively.}
\begin{threeparttable}
\setlength{\tabcolsep}{3pt} 
\renewcommand{\arraystretch}{1.5} 
\begin{tabular}{lcccc}
  \toprule\toprule
  \multirow{2}{*}{\raisebox{-\heavyrulewidth}{Material}} & \multicolumn{4}{c}{Band gap (eV)} \\
  \cmidrule(lr){2-5}
  & SCAN & HSE$\alpha$ & GW & Exp. \\
  \midrule
  MoS$_2$ & 1.82 & 2.49 & 2.40, 2.48, 2.76 & 2.40\cite{gusakova2017electronic}, 2.50\cite{gusakova2017electronic}  \\
  MoSe$_2$ & 1.54 & 2.18 & 2.08, 2.41, 2.18 & 2.15\cite{zhang_probing_2015},2.18\cite{gusakova2017electronic}  \\
  MoTe$_2$ & 1.09 & 1.71 & 1.65, 1.72 & 1.72\cite{yang2015robust}  \\
  WS$_2$ & 1.60 & 2.51 & 2.46, 2.43 & 2.50\cite{Schuler2019}  \\
  WSe$_2$ & 1.34 & 2.13 & 2.01, 2.08 & 2.12\cite{zhang_probing_2015}  \\
  WTe$_2$ & 0.92 & 1.53 & - & -  \\
   \bottomrule\bottomrule
\end{tabular}
\label{table:bandgap}
\end{threeparttable}
\end{center}
\end{table}

The calculated electronic band structure with orbital-resolved contributions for the monolayer TMDs are shown in Fig.~\ref{fig2}. All the monolayer TMDs, except WSe$_2$ and WTe$_2$, have direct band gaps with the valence-band maxima (VBM) and conduction-band minima (CBM) located at the K point in the 2D Brillouin zone, in agreement with previous studies \cite{jin2013direct,pandey2020layer,silva2016electronic,Robert2016}. In the case of WSe$_2$ and WTe$_2$, the CBM occurs at the Q point (midway between $\Gamma$ and K), giving an indirect band gap that is also in agreement with previous results for WSe$_2$\cite{zhang_probing_2015,hsu_evidence_2017}. In the case of WTe$_2$, experimental information on the electronic structure is not available, since this compound is most stable in the 1T' structure \cite{tang_quantum_2017}, not in the 1H structure as studied here.
The VBM and CBM are mostly composed of transition-metal $d$ orbitals with a minor contribution from the chalcogen $p$ orbitals. To be more specific, the highest energy valence bands of the TMDs studied here are composed of $d$\textsubscript{$x$\textsuperscript{2}-$y$\textsuperscript{2}} and $d$\textsubscript{$x$$y$} orbitals of the metal atom and $p$\textsubscript{$x$} and $p$\textsubscript{$y$} orbitals of the chalcogen atoms, whereas the lowest conduction band is mainly composed of $d$\textsubscript{$z$\textsuperscript{2}} orbital of the metal atom with some contribution from $p$\textsubscript{$x$} and $p$\textsubscript{$y$} orbitals of the chalcogen atoms. Note that the splitting due to SOC at the VBM increases by going from S to Te, since Te is heavier than Se and S. The dispersion of the highest valence bands and lowest conduction bands decreases by going from S to Te, which we attribute to the increase in lattice parameters. In other words, as the metal-metal distance increases, the band near the VBM and CBM becomes flatter due to the decrease in the orbital overlap of the metal atoms.

The fundamental band gaps of monolayer TMDs, calculated using SCAN and HSE$\alpha$, are listed in Table~\ref{table:bandgap}. The results using SCAN are in good agreement with previous reports \cite{guo2016band,Lambrecht2012,Robert2016}; however, SCAN underestimates the fundamental band gap by more than $\sim$0.6-0.9 eV. Our results using HSE$\alpha$ are systematically closer to quasi-particle GW calculations \cite{Kim2021,Lambrecht2012,Robert2016,gusakova2017electronic} and the experimental data from scanning tunneling spectroscopy (STS) measurements \cite{zhang_probing_2015,gusakova2017electronic,Zhang2017,Schuler2019}, and are much larger than the optical gaps, as expected due to the large exciton binding energies. The experimental values for the band gap are taken from low-temperature STS measurements\cite{zhang_probing_2015,gusakova2017electronic,Zhang2017,Schuler2019}, where a metal tip is placed on top of the pristine material, and a bias is applied between the tip and a metallic substrate that typically pins the Fermi level in the gap of the semiconducting monolayer TMD. By varying the bias voltage and measuring the current, the VBM and CBM are then determined by the onsets in the dI/dV [or log(dI/dV)] spectra by varying the bias voltage toward negative and positive values. We note that obtaining fundamental band gaps based on photoluminescence measurements is not always straightforward, due to the difficulty in simultaneously determining the excitonic peaks and quasi-particle energies in the same spectrum, and having to rely on calculated exciton binding energies \cite{Ugeda2014,gusakova2017electronic,Zhang2017,Schuler2019}.


\begin{figure}
\includegraphics[width=8.5 cm]{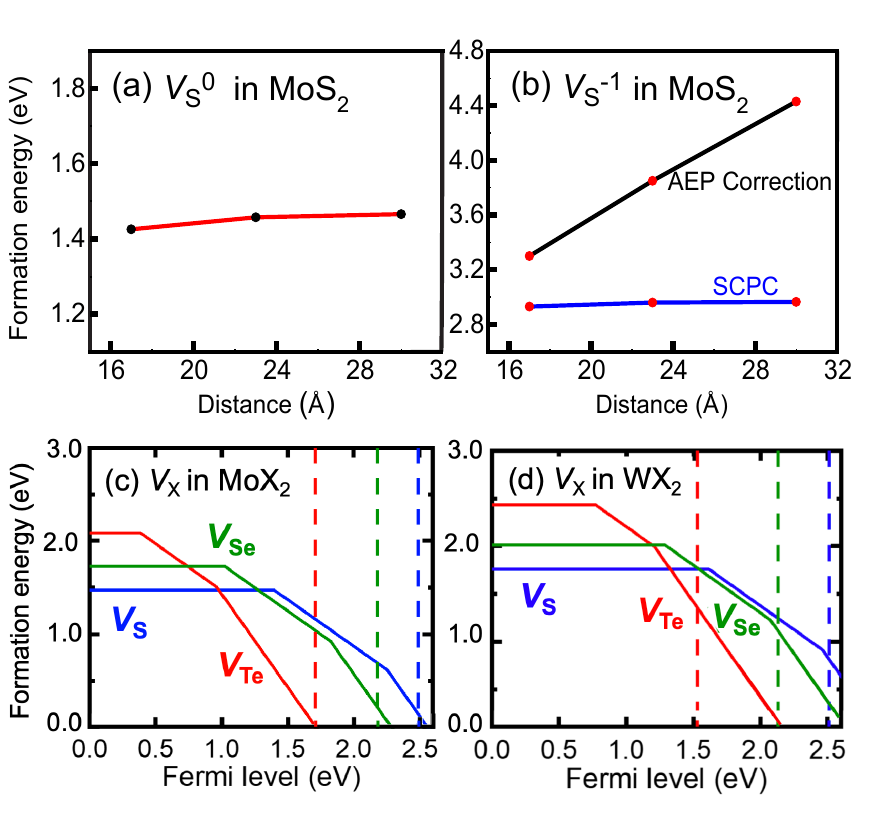}
\caption{Formation energy of the S vacancy in monolayer MoS$_2$ as a function of the "vacuum thickness" separating neigboring monolayers in a 3D periodic structure: (a) neutral and (b) 1$-$ charge state at the valence-band maximum (VBM) in S-poor condition. AEP corresponds to charge state correction based on aligning the average electrostatic potential only, whereas SCPC is the self-consistent charge-state correction from Ref.~\onlinecite{silva2021}. Formation energy as a function of the Fermi energy for chalcogen vacancy in (c) MoX$_2$ and (d) WX$_2$  (X=S, Se, Te) in the chalcogen-poor chemical potential limit. The VBM is used as a reference and the band gap is indicated by dashed lines in each case.}
\label{fig3}
\end{figure}

\begin{figure}
\includegraphics[width=8.5 cm]{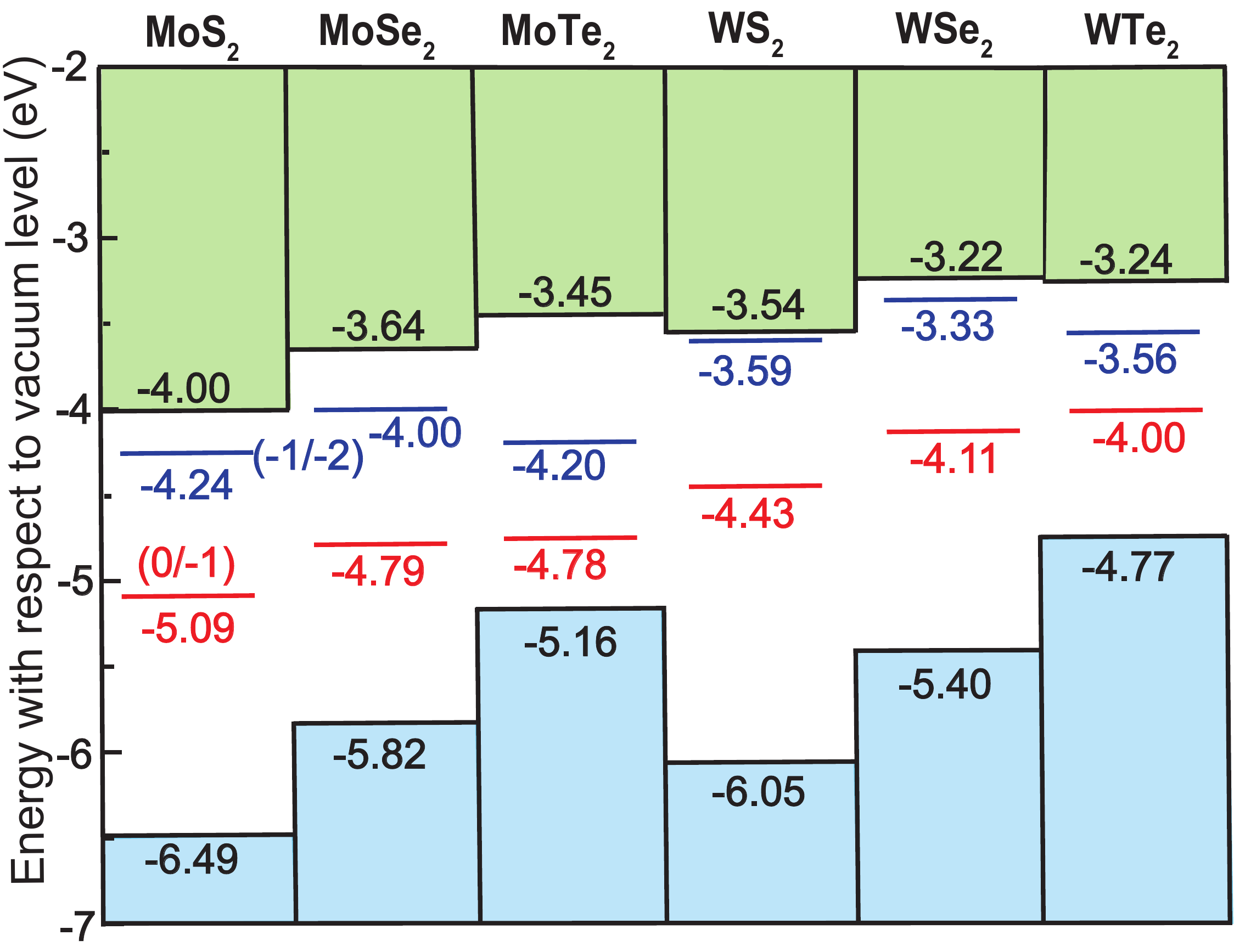}
\caption{Thermodynamic transition levels (0/$-$1) and ($-$1/$-$2) for the chalcogen vacancies in monolayer TMDs.  The positions of the VBM and CBM at the K-point and Q-point (WSe$_2$ and WTe$_2$) in the Brillouin zone with respect to the vacuum level were calculated using HSE$\alpha$.}
\label{fig4}
\end{figure}

The chalcogen vacancy in a given TMD is simulated by removing one chalcogen atom from a supercell containing 90 atoms as shown in Fig.~\ref{fig1}(d). Removing a chalcogen atom leaves three metal-atom dangling bonds that combine into three defect single-particle states.  In the case of the neutral-charge sulfur vacancy in MoS$_2$, the lowest-energy defect state is occupied with two electrons and resonant in the valence band, at 0.17 eV below the VBM; the two other single-particle states are degenerate and unoccupied, located in the upper part of the band gap. Adding one or two electrons to the gap states leads to the negative ($-$1) and doubly negative ($-$2) charge states
of the chalcogen vacancy. The formation energy for the chalcogen vacancy in charge state $q$ is given by\cite{Freysoldt2014}:

\begin{multline*}
E^f[V_X^q]=E_{tot}[V_X^q]-E_{tot}[MX_2]+ E_{tot}[X] + \mu_X 
\\+ q(E_{F}+E_{VBM}) + \Delta^q 
\label{equation:formation}
\end{multline*}
where $E_{tot}[V_X^q]$ is the total energy of the supercell containing the vacancy in charge state $q$, $E_{tot}[MX_2]$ is the total energy of the perfect TMD monolayer using the same supercell, $E_{tot}[X]$ is the total energy per atom of the elemental phase of the chalcogen X, and $\mu_X$ is the chemical potential of the species $X$; $\mu_X$=0 represents the $X$-rich limit condition, whereas $\mu_X=\Delta H^f[MX_2]$/2 represents the $X$-poor limit condition which is the most favorable to form the chalcogen vacancy.

The term $\Delta^q$ is the charge-dependent correction due to the finite size of the supercell, for  
which we used the recently proposed self-consistent potential correction (SCPC) for charged defects in 2D materials \cite{Silva2021prl}. In Fig.~\ref{fig3}, we show how the formation energy of S vacancy in MoS$_2$ changes linearly with the vacuum size if only the correction based on the alignment of the average electrostatic potential (AEP) is accounted for. We also show how the formation energy varies with the vacuum thickness when using the SCPC correction; the method's accuracy is clearly evident. The calculations using the SCPC method show that the formation energy of the negatively charged S vacancy in MoS$_2$ changes by less than 0.05 eV when the vacuum thickness increases from 17 to 30 \AA. In contrast, the correction based only on the alignment of the AEP shows that the formation energy does not converge, as expected. 

The calculated formation energies of the chalcogen vacancies in the TMDs, shown in Fig.~\ref{fig3}(c) and (d), show two thermodynamic charge-state transition levels in the gap (kinks in the formation energy plots).  The vacancies behave as deep acceptors with (0/$-$1) and ($-$1/$-$2) levels in the upper part of the band gap.  In the case of the neutral charge state, the doubly degenerate state in the gap is empty;  in the $-$1 charge state, it is singly occupied, resulting in a Jahn-Teller distortion and a paramagnetic state with S=1/2; in the $-$2 charge state, the doubly degenerate state is occupied by two electrons with parallel spins, forming a paramagnetic S=1 center in a collinear spin representation. 

The (0/$-$1) and ($-$1/$-$2) thermodynamic transition levels of the chalcogen vacancies are shown together with band offsets between the monolayer TMDs in Fig.~\ref{fig4}. The band offsets were obtained by determining the position of the VBM and CBM of each TMD with respect to the vacuum level, representing the ionization potential and electron affinity, respectively. In the case of MoX$_2$, going from S to Te, the VBM decreases (ionization potential increases), and the CBM increases (electron affinity decreases) in an absolute energy scale. The valence-band offsets are much larger than the conduction-band offsets. This follows the trend observed for the valence $p$ atomic orbital energies of Te, Se, and S. For example, the VBM of MoTe$_2$ is higher than the VBM of MoSe$_2$ and MoS$_2$ by 0.66 eV and 1.33 eV, respectively, whereas the CBM is higher by only 0.19 eV and 0.55 eV, in good agreement with previous studies \cite{kang2013band}. The valence band offsets are also larger than the conduction-band offsets in WX$_2$, with the CBM of WTe$_2$ being slightly lower than that on WSe$_2$. 

The fact that the (0/$-$1) and ($-$1/$-$2) levels are closer to each other and closer to the VBM in MoTe$_2$ and WTe$_2$ is attributed to the much larger equilibrium distances between the transition-metal atoms is these compounds.  For either MoX$_2$ or WX$_2$, the larger Mo-Mo or W-W distances in the Te-based compounds implies in less overlap between the dangling bonds that compose the chalcogen vacancy states, resulting is weaker electron-electron repulsion between the electrons occupying the defect state in the band gap.  This effect also explains the trend in the ($-$1/$-$2) levels going from S, to Se and to Te compounds for the same transition-metal atom.




Our calculations also allow us to lend insight into the instances where chalcogen vacancies have been invoked to explain experimental results. Shen \textit{et al.} \cite{shen2022healing} concluded that the S vacancy in MoS$_2$ is a donor, and they suggest that using oxygen annealing will passivate the sulfur vacancies, thus suppressing the source of n-type conductivity. The observed n-type conductivity in WS$_2$ was also attributed to S vacancies \cite{chee2017sulfur}. Guo \textit{et al.} \cite{guo2019electronic} claimed to suppress the electron donor level created by the Se vacancy in MoSe$_2$ with the introduction of a halogen atom. Electron doping in WSe$_2$ was explained by the formation of Se vacancies produced during proton irradiation \cite{wang2021controllable,liang2023robust}. Our calculations show that all the chalcogen vacancies lead to deep acceptor states, with both (0/$-$1) and (1-/2-) transition levels in the band gap, and unequivocally establish that they will not contribute to n-type conductivity. We note that most of the monolayer TMDs show unintentional n-type conductivity, the origin of which is still unknown; it is likely due to impurities instead of chalcogen vacancies. 


MoTe$_2$ has been shown experimentally to exhibit unintentional p-type conductivity \cite{liu2020controlling,chen2017contact,kim2017fermi,zhou2015large,pan2016ultrafast,rani2018control,rani2019tuning}.  Based on our calculations, the Te vacancy leads to an acceptor level 0.38 eV above the valence band, and thus, it is unlikely to be effectively ionized at room temperature. Therefore, we rule out Te vacancies as a cause of p-type conductivity in MoTe$_2$, and the source of the observed p-type conductivity in MoTe$_2$ remains unknown. While chalcogen vacancies in MoS$_2$, MoSe$_2$, WS$_2$, and WSe$_2$ cannot cause n-type conductivity, they can act as compensation centers or ``electron killers.'' In samples where n-type conductivity is observed, filling these vacancies would enhance the conductivity. 

In summary, we show that the fundamental band gaps of monolayer TMDs can be obtained from HSE$\alpha$ calculations, and are in agreement with previous reports based on quasi-particle GW calculations and scanning tunneling spectroscopy measurements. We determine
the properties of chalcogen vacancies in MoS$_2$, MoSe$_2$,  MoTe$_2$, WS$_2$, WSe$_2$ and WTe$_2$ using fundamental (and not optical) band gaps as a reference, and show that all chalcogen vacancies act as deep acceptors. All chalcogen vacancies have two acceptor transition levels [(0/$-$1) and ($-$1/$-$2)] that reside in the gap, and lead to S=1/2 and S=1 paramagnetic configurations. By establishing chalcogen vacancies as deep acceptors in each of these monolayer TMDs, we conclude that they cannot account for the frequently reported n-type or p-type conductivity.

\section*{Acknowledgement}

This work was supported by the Laboratory Directed Research and Development (LDRD) Program (Grant No. PPPL-132) at Princeton Plasma Physics Laboratory under U.S. Department of Energy Prime Contract No. DE-AC02-09CH11466. The United States Government retains a non-exclusive, paid-up, irrevocable, world-wide license to publish or reproduce the published form of this manuscript, or allow others to do so, for United States Government purposes. A.J. acknowledges funding from the National Science Foundation (NSF) award \#OIA-2217786. The calculations were carried out at the National Energy Research Scientific Computing Center (NERSC), a U.S. Department of Energy Office of Science User Facility located at Lawrence Berkeley National Laboratory, operated under Contract No. DE-AC02-05CH11231 using NERSC award BES-ERCAP20424, the Stellar Cluster at Princeton University, and the DARWIN computing system at the University of Delaware, using the NSF grant no.~1919839. DW and JLL acknowledge support from the ONR/NRL 6.1 Base Research Program.

\bibliography{monolayer_TMD}

\end{document}